\documentclass[letterpaper,titlepage,11pt]{article}
\usepackage{amsmath,amssymb,mathtools,bm}
\usepackage{physics}
\usepackage{cancel}
\usepackage{slashed}
\usepackage{mathrsfs}
\usepackage{color}
\usepackage[svgnames]{xcolor}
\usepackage{enumerate}
\usepackage{graphicx}
\usepackage{tikz}
\usepackage{caption}
\usepackage[symbol]{footmisc}
\usepackage{mathtools}
\usepackage{enumitem}

\usepackage{multirow}
\usepackage{booktabs}
\usepackage{siunitx}

\usepackage{physics}

\usepackage{tikz}
\usetikzlibrary{arrows.meta, positioning, calc, decorations.pathmorphing, shapes.geometric}
\usepackage{tikzsymbols}
\usepackage{fontawesome5}
\usepackage{xcolor,pifont}
\usetikzlibrary{shadows}

\usepackage[numbers,sort&compress]{natbib}
\allowdisplaybreaks
\usepackage[
colorlinks=true,
linkcolor=DarkBlue,
anchorcolor=Black,
urlcolor=DarkRed,
filecolor=green,
citecolor=DarkRed,
linktoc=page,
pdfstartview=FitV,
pdftitle={},
pdfauthor={Omer Guleryuz},
pdfsubject={},
pdfkeywords={},
bookmarksopen=true
]{hyperref}

\usepackage[nameinlink,capitalise]{cleveref}

\newcommand{\mpl}{M_{\rm Pl}}
\newcommand{\GB}{\mathcal G}
\newcommand{\calV}{\mathcal V}
\newcommand{\ns}{n_s}
\newcommand{\rten}{r}
\newcommand{\As}{A_s}
\newcommand{\alphas}{\alpha_s}

\renewcommand{\dd}{\mathrm d}
\renewcommand{\order}{\mathcal O}
\newcommand{\eff}{\mathrm{eff}}

\newcommand{\EulerChar}{\chi_{\rm top}}

\begin{document}

    \begin{titlepage}
			%%%%%%%%%%%%%%%%%%%%%%%%%%%%%%%%%%
			
			%%%%%%%%%%%%%%%%%%%%%%%%%%%%%%%%%%
			\bigskip
			
			\begin{center}
				{\LARGE\bfseries String Thresholds, Dynamical Gauss--Bonnet Couplings, and Starobinsky Attractors}
				\\[10mm]
				\textbf{Omer Guleryuz}\\[5mm]
				\vskip 25pt

				{\em  \hskip -.1truecm Department of Physics, Istanbul Technical University,  \\
					Maslak 34469 Istanbul, Türkiye  \vskip 5pt }

				{{\tt \href{mailto:omerguleryuz@itu.edu.tr}{omerguleryuz@itu.edu.tr}}
    
    }

			\end{center}
			
			\vspace{3ex}

			\begin{center}
				{\bfseries Abstract}
			\end{center}
			\begin{quotation}

We develop a string-motivated dynamical Gauss--Bonnet completion of Starobinsky inflation. Since a constant Gauss--Bonnet term is topological in four dimensions, observable effects must arise from a modulus, dilaton, or compactification threshold whose value changes during inflation. We formulate the system as a scalar--Gauss--Bonnet effective theory, derive an invariant matching between the threshold-corrected plateau and the leading CMB observables \(\ns\), \(\rten\), and the running \(\alphas\equiv\dd\ns/\dd\ln k\), and impose explicit string and Kaluza--Klein cutoff bounds. Calabi--Yau topology and string threshold amplitudes are used only as microscopic priors for the threshold function; the observable deformation is fixed only after stabilization, trajectory selection, and single-clock matching. In the controlled heavy-modulus regime, a positive matched deformation raises the scalar tilt, lowers the tensor signal, and makes the running mildly less negative. A representative \(X_{24}(1,1,2,8,12)\) example illustrates how topological data and an effective threshold response define a quantitative compactification target for the range \(\kappa_G\simeq7\text{--}17\), while emphasizing that this is not a direct prediction from a fully stabilized compactification.
			\end{quotation}
			
			\vfill
			
			%%%%%%%
            \begin{center}{\flushleft{\today}}\end{center}
			%%%%%%
		\end{titlepage}
		\setcounter{page}{1}
		\tableofcontents

		%%%%%%%%%%%%%%%%%%%%%%%%%%%%%%%%%%%%%%%%%%%%%%%%%%%%%%%%%%%%%%%%%
\newpage

\section{Introduction}
\label{sec:introduction}
Starobinsky inflation is the canonical curvature-squared plateau model. In the Einstein frame, it is equivalent to a scalaron rolling on an exponentially approached plateau, and it predicts a small tensor-to-scalar ratio compatible with current CMB limits
\cite{Starobinsky:1980te,Mijic:1986iv,Whitt:1984pd,Maeda:1988ab,DeFelice:2010aj,Planck:2018vyg,Planck:2018jri,BICEP:2021xfz}.
Its leading predictions,
\begin{equation}
    \ns \simeq 1-\frac{2}{N_*}, \qquad \rten \simeq \frac{12}{N_*^2},
\end{equation}
where \(N_*\) is the number of e-folds before the end of inflation, make it a robust benchmark for controlled deformations of plateau inflation.

Current data make small, theoretically organized departures from this attractor worth studying. Planck remains compatible with the canonical prediction, while ACT DR6 and recent CMB+BAO combinations provide useful higher-\(\ns\) target values \cite{Planck:2018jri,AtacamaCosmologyTelescope:2025blo,AtacamaCosmologyTelescope:2025nti,Balkenhol:2025wms}. This should not be interpreted as a discovery-level anomaly. It is, instead, a motivation to ask whether a controlled ultraviolet correction can modestly raise \(\ns\) while preserving the tensor safety and effective-field-theory control of the plateau.

String effective actions provide a natural setting for this question. At leading order in the inverse string tension, the Gauss--Bonnet combination is distinguished among curvature-squared invariants because it avoids the massive spin-2 ghost associated with generic Ricci-squared or Riemann-squared corrections \cite{Zwiebach:1985uq,Callan:1985ia,Gross:1986mw}. In four dimensions, however, a constant Gauss--Bonnet term is topological. It cannot modify the local inflationary background, perturbations, or observables. The conservative four-dimensional setup is therefore not
\[R+\alpha R^2+\xi\GB\]
with constant \(\xi\), but a scalar-dependent threshold coupling,
\begin{equation}
    R+\alpha R^2+\xi\GB
    \quad \longrightarrow\quad
    R+\alpha R^2+f(\psi)\GB,
\end{equation}
where \(\psi\) denotes a modulus, dilaton, or effective compactification threshold field.

In four dimensions, a scalar-dependent Gauss--Bonnet coupling is equivalent, up to integrations by parts, to a particular Horndeski/generalized Galileon scalar--tensor interaction, and it yields second-order field equations despite the curvature-squared structure
\cite{Horndeski:1974wa,Deffayet:2011gz,Kobayashi:2011nu,Kanti:1995vq,Nojiri:2005vv,Guo:2009uk}.
Compactification data and string-threshold amplitudes can set priors on the sign, scale, and functional form of \(f(\psi)\). In particular, Calabi--Yau topology supplies discrete topological weights, while one-loop and BPS-saturated threshold calculations motivate moduli-dependent gravitational couplings. These data do not directly predict CMB observables: the observables arise only after the threshold-active field is stabilized and its accepted trajectory is matched to a single-clock plateau.

This paper develops the corresponding effective field theory and its leading observational consequences.\footnote{This manuscript supersedes the withdrawn preprint arXiv:1808.06404. Only the restricted Horndeski projection described in \cref{sec:galileonInterpretation} is retained; the earlier phenomenological slow-roll ansatz and formulae are not used.} The central result is a two-step matching. Compactification thresholds and heavy-modulus stabilization determine invariant coefficients in the large-field scalaron plateau; the attractor expansion then maps those coefficients to CMB observables. The microscopic data enter through a local threshold response, while the observable deformation is encoded in a matched infrared parameter \(\kappa_G\), defined from the invariant plateau coefficient in \cref{sec:derivation}.
With \(\alphas\equiv \dd\ns/\dd\ln k\), the leading attractor map is
\begin{equation}
    \Delta \ns \simeq \frac{2\kappa_G}{N_*^2}, \qquad \rten \simeq \frac{12}{N_*^2} \left(1-\frac{\kappa_G}{N_*}\right), \qquad \alphas\simeq-\frac{2}{N_*^2}+\frac{4\kappa_G}{N_*^3},
\end{equation}
up to higher-order terms in the attractor expansion and assuming that \(\kappa_G\) varies slowly over the CMB window. The deformation should not be viewed as a commitment to any particular high-\(\ns\) data set: the Starobinsky point is recovered continuously as \(\kappa_G\to0\). Positive \(\kappa_G\) is useful because it raises the scalar tilt while lowering the tensor signal, so the tensor prediction remains safely below current bounds throughout the benchmark range. The running is included as a correlated consistency observable; in the benchmark regime, its correction remains at the \(10^{-4}\)--\(10^{-3}\) level and is not the primary discovery channel.

We also give a semi-microscopic target diagnostic based on the \(X_{24}(1,1,2,8,12)\) topological weight. This example is not a full compactification prediction. Rather, it shows how known string-threshold structures, Calabi--Yau topology, and an effective response coefficient can be organized into a quantitative target for the phenomenologically relevant interval \(\kappa_G\simeq7\text{--}17\). A genuine prediction would require computing the threshold function, the stabilized modulus mass matrix, and the accepted single-clock trajectory in a single explicit compactification.

The paper is organized as follows. \Cref{sec:eftbaseline} fixes the Starobinsky and higher-curvature EFT baseline and states the four-dimensional Gauss--Bonnet obstruction. \Cref{sec:stringcy} summarizes the compactification threshold priors and admissible trajectories. \Cref{sec:dynamicalanalysis} derives the matched observables and validity conditions. \Cref{sec:phenomenology} gives benchmark comparisons with current data and develops a semi-microscopic threshold benchmark connecting representative Calabi--Yau data to the target \(\kappa_G\) range. \Cref{sec:discussionconclusions} closes with the interpretation and conclusions.

\section{Starobinsky and higher-curvature EFT constraints}
\label{sec:eftbaseline}
This section fixes the scalaron normalization, separates ordinary higher-curvature deformations from the four-dimensional Gauss--Bonnet obstruction, and states the scalar--tensor threshold interpretation used below.

\subsection{Starobinsky baseline and nearby deformations}
\label{sec:starobinsky}
We work in reduced Planck units, \(\mpl=1\), unless dimensions are displayed explicitly. The Starobinsky action is
\begin{equation}
S_{R^2}=\frac12\int \dd^4x\sqrt{-g}\left(R+\frac{R^2}{6M^2}\right),
\end{equation}
which is equivalent to a canonical scalaron with
\begin{equation}
U_S(\varphi)=\frac{3M^2}{4} \left(1-e^{-\sqrt{2/3}\varphi}\right)^2.
\label{eq:staropot}
\end{equation}
The standard large-field formulae, reviewed in \cref{app:starobinskyBaseline}, are
\begin{equation}
\ns^{(0)}=1-\frac2N+\order\!\left(\frac{\ln N}{N^2}\right), \qquad \rten^{(0)}=\frac{12}{N^2}+\order\!\left(\frac{\ln N}{N^3}\right), \qquad \alphas^{(0)}=-\frac{2}{N^2}+\order\!\left(\frac{\ln N}{N^3}\right).
\end{equation}
The auxiliary-field derivation also clarifies a common but incorrect shortcut involving the Gauss--Bonnet term. In controlled \(f(R)\) or algebraic nonminimal sectors, Hubbard--Stratonovich or Legendre linearization, algebraic heavy-field elimination, and the Jordan--Einstein Weyl map can be organized consistently; this is the frame-safe logic behind the scalaron EFT and the robustness of the \(R^2\) baseline
\cite{Arapoglu:2025hpr}. In a general \(f(R)\) theory with gravitational Lagrangian \(\frac12 f(R)\), we denote derivatives with respect to \(R\) by
\begin{equation}
f_{,R}\equiv\frac{\partial f}{\partial R}, \qquad f_{,RR}\equiv\frac{\partial^2 f}{\partial R^2}.
\end{equation}
When \(f_{,R}>0\), the Einstein-frame metric is obtained by the Weyl rescaling
\begin{equation}
g^E_{\mu\nu} = \Omega^2 g^J_{\mu\nu}, \qquad \Omega^2=f_{,R}.
\end{equation}
For the Starobinsky choice \(f(R)=R+R^2/(6M^2)\), this gives
\begin{equation}
\Omega^2 = 1+\frac{R}{3M^2} = e^{\sqrt{2/3}\varphi}.
\end{equation}
This construction works because the relevant curvature dependence is captured by an auxiliary scalar or an algebraic nonminimal coupling. It cannot be applied to \(R+\alpha R^2+\xi\GB\) by treating the independent invariant \(\GB\) as if it were a function of \(R\). Thus, it is not legitimate to define a would-be Weyl factor by
\begin{equation}
    \Omega_{\rm bad}^2 \stackrel{?}{=} \frac{\partial}{\partial R} \left( R+\alpha R^2+\xi\GB \right)
\end{equation}
and then replace \(\GB\) by its quasi-de Sitter background value. The \(R^2\) sector is treated with the standard \(f(R)\) auxiliary/Weyl machinery; the \(f(\psi)\GB\) term is kept as a genuine scalar--tensor threshold and matched only after the stabilized trajectory is known.
\begin{table}[htb]
\centering
\caption{Simple deformations of the Starobinsky action and their effective-field-theory status.}
\begin{tabular}{llll}
\toprule
Deformation & Propagating content & Main issue & Role in this work\\
\midrule
\(R^2\) & scalaron & none in EFT regime & baseline\\
\(R^3\) & scalaron only & plateau control and tuning & comparison\\
\(R_{\mu\nu}R^{\mu\nu}\) & massive spin-2 mode & ghost & discarded\\
constant \(\xi\GB\) & no local 4D mode & topological & replaced\\
\(f(\psi)\GB\) & scalar--tensor sector & stability and modulus control & main model\\
\bottomrule
\end{tabular}
\label{tab:simpledeformations}
\end{table}
\Cref{tab:simpledeformations} separates the cases relevant for this paper. Pure \(f(R)\) deformations such as \(R^3\) keep only the scalaron but give model-dependent plateau corrections \cite{Aldabergenov:2018qhs,Cheong:2020rao}; the perturbative \(R^3\) expansion is collected in \cref{app:higherCurvatureComparison}. Generic Ricci-squared or Weyl-squared corrections instead carry a massive spin-2 ghost \cite{Stelle:1976gc,Stelle:1977ry,Woodard:2015zca}. A constant four-dimensional \(\xi\GB\) term avoids that ghost but is topological, so the locally dynamical route is the scalar-dependent threshold \(f(\psi)\GB\).

\subsection{Four-dimensional Gauss--Bonnet obstruction}
\label{sec:topology}
The four-dimensional Gauss--Bonnet density is
\begin{equation}
\GB = R^2 -4R_{\mu\nu}R^{\mu\nu} +R_{\mu\nu\rho\sigma}R^{\mu\nu\rho\sigma}.
\end{equation}
Its integral is topological. More precisely, on a compact four-manifold without boundary,
\begin{equation}
\int \dd^4x\sqrt{-g}\,\GB = 32\pi^2\,\EulerChar(M_4),
\end{equation}
while on a spacetime with boundary, its variation reduces to a boundary contribution after the appropriate Gauss--Bonnet boundary term is included. Hence, a constant coefficient multiplying \(\GB\) cannot modify the local four-dimensional field equations:
\begin{equation}
\delta\int \dd^4x\sqrt{-g}\,\xi\GB=0, \qquad \xi=\mathrm{constant},
\end{equation}
up to boundary data \cite{Lovelock:1971yv,Zwiebach:1985uq}. This elementary fact rules out treating
\begin{equation}
R+\alpha R^2+\xi\GB
\end{equation}
as a locally modified four-dimensional inflationary model when \(\xi\) is constant.

This gives the central four-dimensional obstruction: a constant Gauss--Bonnet coefficient cannot alter the local inflationary background or perturbation equations. Any correction to \(\ns\), \(\rten\), \(\alphas\), or the scalar and tensor propagation speeds must therefore come from a dynamical coupling, extra-dimensional physics, boundary data, or an equivalent scalar--tensor completion of a limiting procedure.

This distinction is important in view of the influential four-dimensional Einstein--Gauss--Bonnet proposal of Glavan and Lin, which starts from the \(D\)-dimensional Lovelock term, rescales the Gauss--Bonnet coupling by \((D-4)^{-1}\), and only then considers a \(D\to4\) limit \cite{Glavan:2019inb}. That prescription motivated a large literature, but later analyses clarified that nontrivial four-dimensional dynamics require a specified limiting or regularization procedure, or an equivalent scalar--tensor description \cite{Gurses:2020ofy,Lu:2020iav}. The present paper does not use the bare rescaled \(D\to4\) prescription as an inflationary model. It keeps the ordinary four-dimensional topological obstruction intact and introduces local dynamics only through the scalar-dependent threshold \(f(\psi)\GB\).

The minimal two-field effective action used below is
\begin{equation}
S = \int \dd^4x\sqrt{-g} \left[\frac12 R -\frac12(\partial\varphi)^2 -\frac12(\partial\psi)^2-U(\varphi,\psi)-f(\psi)\GB\right].
\label{eq:twofieldaction}
\end{equation}
Here \(\varphi\) denotes the Starobinsky scalaron and \(\psi\) denotes the modulus, dilaton, or other compactification field that carries the Gauss--Bonnet threshold. The sign convention in front of \(f(\psi)\GB\) is not physical by itself; it can be absorbed into \(f\to -f\). What is physical is the field dependence of \(f\), and in particular, the derivatives of \(f\) that enter the background and perturbation equations.

\subsection{Horndeski/Galileon interpretation of the threshold coupling}
\label{sec:galileonInterpretation}
The action in \cref{eq:twofieldaction} also gives the precise relation to the
Galileon/G-inflation language. Define
\begin{equation}
X_\psi=-\frac12\nabla_\mu\psi\nabla^\mu\psi .
\end{equation}
For a term \(+\xi(\psi)\GB\), one convenient Horndeski representation is, up to total
derivatives,
\begin{equation}
\sqrt{-g}\,\xi(\psi)\GB \doteq \sqrt{-g}\sum_{i=2}^5\mathcal L_i[G_i^{\rm GB}],
\label{eq:gbHorndeskiMap}
\end{equation}
where \(\doteq\) denotes equality modulo boundary terms and, in the common convention
\(\mathcal L_3=-G_3\Box\psi\),
\begin{align}
G_2^{\rm GB} &= 8\xi_{,\psi\psi\psi\psi}X_\psi^2 \left(3-\ln\frac{X_\psi}{\mu^4}\right), \nonumber\\ G_3^{\rm GB}
&=4\xi_{,\psi\psi\psi}X_\psi \left(7-3\ln\frac{X_\psi}{\mu^4}\right), \nonumber\\
G_4^{\rm GB}
&=
4\xi_{,\psi\psi}X_\psi \left(2-\ln\frac{X_\psi}{\mu^4}\right), \nonumber\\ G_5^{\rm GB}
&= -4\xi_{,\psi}\ln\frac{X_\psi}{\mu^4}.
\label{eq:gbHorndeskiFunctions}
\end{align}
The logarithm appears only in this particular Horndeski representative of the
scalar--Gauss--Bonnet interaction. It should not be interpreted as a physical singularity of the original \( \xi(\psi)\GB \) theory as \(X_\psi\to0\): for smooth \(\xi\), the original scalar--Gauss--Bonnet coupling is regular, and the local background equations depend on \(\xi_{,\psi}\), \(\dot\xi\), and \(\ddot\xi\), not on an independent logarithmic potential. The arbitrary scale \(\mu\) labels the representative form. Changing \(\mu\) changes only the representative Horndeski functions, leaving the original scalar--Gauss--Bonnet interaction equivalent up to integrations by parts. For the sign convention of \cref{eq:twofieldaction},
one sets \(\xi=-f\). The important point is not the particular representative in
\cref{eq:gbHorndeskiFunctions}, but the fact that a scalar--Gauss--Bonnet threshold is a correlated point in Horndeski function space, rather than a generic arbitrary Galileon ansatz \cite{Deffayet:2011gz,Kobayashi:2011nu}.

This is the controlled sense in which the earlier G-inflation intuition enters the present work. The phenomenological G-inflation or kinetic-braiding sector is usually written schematically as
\begin{equation}
\mathcal L_\psi = K(\psi,X_\psi)-G_3(\psi,X_\psi)\Box\psi
\end{equation}
possibly supplemented by nonminimal curvature couplings \cite{Kobayashi:2010cm,Deffayet:2010qz}. The scalar--Gauss--Bonnet threshold contains
braiding-like terms of this type, but also the correlated \(G_4\) and \(G_5\) curvature-derivative operators required by \cref{eq:gbHorndeskiMap}. After the heavy threshold field follows an accepted single-clock trajectory, \(\psi=\psi_0(\varphi)\), one has
\begin{equation}
\nabla_\mu\psi=\psi_0'(\varphi)\nabla_\mu\varphi, \qquad X_\psi=(\psi_0')^2X_\varphi,
\qquad \Box\psi=\psi_0'\Box\varphi-2\psi_0''X_\varphi,
\end{equation}
so the Horndeski operators in \cref{eq:gbHorndeskiMap} project onto induced derivative
self-interactions for the adiabatic scalar. Their coefficients are not free Galileon data: they are tied to the threshold function, the stabilized trajectory, and the subsequent single-clock matching. This projection is valid only in the weak scalar--Gauss--Bonnet, heavy-entropic, small-turn regime. Denoting the entropic mass by \(M_{\rm ent}\) and the covariant turn rate by \(\Omega_{\rm turn}\), this means schematically
\begin{equation}
|H\dot f_{\eff}|\ll1, \qquad \frac{M_{\rm ent}^2}{H^2}\gg1, \qquad \frac{\Omega_{\rm turn}^2}{M_{\rm ent}^2}\ll1.
\end{equation}
This is the same regime in which the invariant single-clock matching to \(A_G\) and
\(\kappa_G\) is used below. In this regime, the leading observational effect is captured by the matched plateau coefficient \(A_G\), or equivalently by \(\kappa_G\). Outside it, the scalar and tensor spectra must be computed from the full scalar--Gauss--Bonnet/Horndeski perturbation system. Thus, the Galileon/G-inflation language is recovered only as a low-energy description of induced Horndeski operators; it is not an independent phenomenological ansatz.

\section{Compactification inputs and threshold trajectories}
\label{sec:stringcy}
This section records the compactification information actually used in the effective theory: the string and Kaluza--Klein cutoff hierarchy, the topological and threshold priors, and the bounded threshold trajectories that can feed the plateau matching.

String effective actions naturally generate higher-curvature corrections. In the string frame, the ten-dimensional heterotic action may be written schematically as
\begin{equation}
S_{10}^{(s)} = \frac{1}{2\kappa_{10}^2} \int \dd^{10}x\sqrt{-g_{10}^{(s)}}\,e^{-2\Phi}
\left[ R_{10}^{(s)} +4(\partial\Phi)^2 +\frac{\alpha'}{8}\GB_{10}^{(s)} +\cdots \right],
\label{eq:tenDstring}
\end{equation}
where \(2\kappa_{10}^2=(2\pi)^7\alpha'^4\) in a common convention. At order
\(\alpha'\), generic curvature-squared terms can be reshuffled by field redefinitions, and one may choose the ghost-free Gauss--Bonnet scheme. In this scheme, the curvature-squared correction is written in Lovelock form, avoiding the massive spin-2 ghost that would be present for a generic quadratic curvature combination
\cite{Callan:1985ia,Zwiebach:1985uq,Gross:1986mw,Antoniadis:1993jc}.

We do not claim a single fully specified compactification containing all ingredients below. The ten-dimensional expression illustrates the origin of Gauss--Bonnet-type curvature-squared schemes, while Calabi--Yau topology and moduli stabilization supply effective priors for four-dimensional threshold functions. The use of these ingredients in this paper is therefore as a threshold-matching framework.

\subsection{String threshold and cutoff bounds}
\label{subsec:normalization}
Let the internal space be a Calabi--Yau threefold \(K_6\) with dimensionless string-frame volume
\begin{equation}
\calV_s = \frac{1}{(2\pi\sqrt{\alpha'})^6} \int_{K_6}\dd^6y\sqrt{g_6^{(s)}}.
\end{equation}
For slowly varying dilaton and volume moduli, dimensional reduction of the Einstein--Hilbert term gives
\begin{equation}
\frac{M_{\rm Pl}^2}{2} = \frac{1}{2\kappa_{10}^2}e^{-2\Phi_0}V_6, \qquad V_6=(2\pi\sqrt{\alpha'})^6\calV_s.
\end{equation}
Equivalently,
\begin{equation}
M_{\rm Pl}^2 = C_{\rm Pl}\frac{\calV_s}{g_s^2\,\alpha'}, \qquad g_s=e^{\Phi_0},
\end{equation}
where \(C_{\rm Pl}\) collects the convention-dependent factors of \(2\pi\) and the reduced Planck-mass normalization. In a common phenomenological convention \(C_{\rm Pl}=1/(2\pi)\). Changing this convention rescales the definition of the dimensionless threshold coefficient but does not affect the local statement that only field dependence in the four-dimensional Gauss--Bonnet threshold is dynamical.

With \(M_s=\alpha'^{-1/2}\), the same normalization gives
\begin{equation}
\frac{M_s}{M_{\rm Pl}} = \frac{g_s}{\sqrt{C_{\rm Pl}\calV_s}}, \qquad \frac{M_{\rm KK}}{M_{\rm Pl}} \sim \frac{g_s}{\sqrt{C_{\rm Pl}}\calV_s^{2/3}},
\end{equation}
where the second relation is the isotropic Kaluza--Klein estimate. The inflationary scale is fixed by
\begin{equation}
\frac{H_*^2}{M_{\rm Pl}^2} = \frac{\pi^2}{2}\As\rten.
\end{equation}
Requiring \(H_*\) to lie parametrically below the string and Kaluza--Klein thresholds gives
\begin{align}
\calV_s &\ll \frac{2g_s^2}{\pi^2 C_{\rm Pl}\As\rten},
\label{eq:stringCutoffBound}
\\
\calV_s &\ll
\left( \frac{2g_s^2}{\pi^2 C_{\rm Pl}\As\rten} \right)^{3/4}.
\label{eq:kkCutoffBound}
\end{align}
The Kaluza--Klein condition is usually the stronger one.

The part of \cref{eq:tenDstring} proportional to the external four-dimensional Gauss--Bonnet density reduces schematically to
\begin{equation}
S_{\GB,4}^{(s)} \supset \int\dd^4x\sqrt{-g_4^{(s)}}\, F_s(\Phi,\calV_s,\psi_i)\,\GB_4^{(s)},
\end{equation}
where \(\psi_i\) denotes the compactification moduli. After passing to the four-dimensional Einstein frame, the constant part of this coefficient remains topological. The locally relevant piece is the field dependence,
\begin{equation}
\int \dd^4x\sqrt{-g_E}\, f(\psi_i)\GB_E, \qquad \partial_i f\neq0.
\end{equation}
This is the four-dimensional string-derived Gauss--Bonnet threshold used below.

\subsection{Topological and threshold priors}
Calabi--Yau compactifications are natural because they preserve four-dimensional supersymmetry and supply K\"ahler and complex-structure moduli \cite{Candelas:1985en,Yau:1978cfy,Denef:2007pq}. For the present threshold analysis, the coarse topological datum used explicitly is the Euler characteristic,
\begin{equation}
\EulerChar(K_6)
= 2\left(h^{1,1}-h^{2,1}\right).
\end{equation}
We package it into the dimensionless weight
\begin{equation}
\mathcal T(K_6)\equiv \frac{\EulerChar(K_6)}{24}.
\label{eq:topologicalPrior}
\end{equation}
This weight is not an observable. It is an input to a threshold prior, to be combined with the moduli dependence and the accepted inflationary trajectory. String compactifications generate moduli-dependent threshold corrections to four-dimensional
gauge and gravitational couplings. In heterotic compactifications, these threshold functions are computed from one-loop amplitudes and depend on the moduli of the internal space \cite{Dixon:1990pc,Antoniadis:1992sa}. In type-II/heterotic \(N=2\) dual pairs, the same higher-derivative gravitational couplings are closely related to genus-one topological-string data and BPS-saturated threshold amplitudes
\cite{Bershadsky:1993cx,Antoniadis:1995zn,Harvey:1995fq}. These results provide the microscopic motivation for treating the four-dimensional Gauss--Bonnet coefficient as a moduli-dependent threshold function.

We use the following schematic threshold prior:
\begin{equation}
f(\psi_i) = f_0+\lambda_{\rm GB}\,\mathcal F(\psi_i), \qquad \lambda_{\rm GB} \sim c_{\rm GB}\, \mathcal P(g_s,\calV_s,\alpha')\, \mathcal T(K_6).
\label{eq:lambdaGeneral}
\end{equation}
Here \(\mathcal F\) is a dimensionless threshold-shape function, \(f_0\) is locally irrelevant, \(\mathcal P\) encodes the conversion to four-dimensional Einstein-frame units, and \(\mathcal T(K_6)\) is the topological weight in \cref{eq:topologicalPrior}. The symbol \(\sim\) is intentional: fluxes, warping, bundle data, normalization conventions, and the choice of local moduli patch can all modify the continuous coefficient multiplying the topological weight. Existing string threshold calculations can determine the moduli dependence of \(\mathcal F\), or equivalently, its local derivatives along a chosen region of moduli space. They do not by themselves determine the CMB observables, because the latter also depend on moduli stabilization, the inflationary trajectory, and the single-clock matching.

For the present purpose, the relevant local data are
\begin{equation}
\mathcal F(\psi) = \mathcal F_\infty + \mathcal F_1(\psi-\psi_\infty) + \frac12\mathcal F_2(\psi-\psi_\infty)^2 +\cdots,
\end{equation}
where \(\psi_\infty\) is the endpoint approached by the stabilized modulus in the large-field regime. The constant piece \(f_0+\lambda_{\rm GB}\mathcal F_\infty\) multiplies a topological four-dimensional invariant and does not affect the local equations. The derivatives \(\mathcal F_1,\mathcal F_2,\ldots\), together with the stabilized trajectory, are the quantities that enter the matched plateau deformation. Thus, the observable deformation is not fixed by \(\lambda_{\rm GB}\) or by \(\mathcal T(K_6)\) alone; it is fixed only after solving the threshold-active dynamics and extracting the invariant coefficient \(A_G\) in \cref{sec:derivation}.
\begin{table}[htb]
\centering
\caption{Representative compactification data interpreted as priors for the dynamical
Gauss--Bonnet threshold. The column \(\mathcal T=\EulerChar/24\) gives the topological weight appearing in \cref{eq:lambdaGeneral}; it is not an observable prediction.}
\begin{tabular}{lcccc}
\toprule
Example & \(h^{1,1}\) & \(h^{2,1}\) & \(\EulerChar(K_6)\) & \(\mathcal T=\EulerChar/24\) \\
\midrule
Quintic hypersurface \cite{Candelas:1985en} & 1 & 101 & \(-200\) & \(-8.33\) \\
\(\mathbb P^4_{1,1,2,2,6}[12]\) benchmark \cite{Kachru:1995wm,Klemm:1995tj} & 2 & 128 & \(-252\) & \(-10.5\) \\
\(X_{24}(1,1,2,8,12)\) / K3-fibered benchmark \cite{Klemm:1995tj} & 3 & 243 & \(-480\) & \(-20\) \\
Enriques / self-mirror benchmark \cite{Ferrara:1995yx} & 11 & 11 & \(0\) & \(0\) \\
\bottomrule
\end{tabular}
\label{tab:cyprior}
\end{table}
The entries in \cref{tab:cyprior} are not meant to define a unique compactification model for the inflationary sector. They provide representative topological weights. The quintic gives the standard one-parameter hypersurface benchmark; the \((2,128)\) entry is a common two-modulus weighted-hypersurface benchmark; the \((3,243)\) entry is the familiar \(X_{24}(1,1,2,8,12)\) K3-fibered benchmark; and the \((11,11)\) entry represents a self-mirror case with no Euler-characteristic enhancement. If the same stabilization sector and the same threshold-shape function \(\mathcal F\) were compared across different compactification topologies, the sign and relative magnitude of the induced threshold force would scale with \(\mathcal T(K_6)\). The resulting \(\kappa_G\), however, is obtained only after solving the stabilized trajectory and extracting the matched plateau coefficients.

\subsection{Admissible threshold trajectories}
\label{subsec:boundedmoduli}
The threshold prior becomes predictive only after the threshold-active modulus follows a controlled large-field trajectory. The threshold field is not assumed to be a second light slow-roll degree of freedom. Rather, it is a heavy field whose instantaneous minimum is displaced by the scalaron-dependent background and by the Gauss--Bonnet threshold force. Its motion is therefore an adiabatic settling of a heavy valley, not an independent multi-field slow-roll trajectory. A useful intrinsic criterion for such controlled motion is bounded finite-endpoint behavior. More generally one may write
\begin{equation}
\psi_0(\varphi) = \psi_\infty+A e^{-\Delta\varphi} +B e^{-2\Delta\varphi}+\cdots = \psi_\infty+A x^{\Delta/a} +B x^{2\Delta/a}+\cdots, \qquad a=\sqrt{\frac23}.
\end{equation}
with \(\Delta>0\). The attractor benchmark focuses on the aligned case \(\Delta=a\),
\begin{equation}
\psi_0(\varphi) = \psi_\infty+A x+B x^2+\order(x^3), \qquad x=e^{-a\varphi}.
\label{eq:admissiblePsi}
\end{equation}
The endpoint \(\psi_\infty\) denotes the stabilized value approached in the large-field regime. In the simplest plateau matching, this endpoint may be identified with the de Sitter-shifted solution \(\bar\psi\) used in the local extraction of Appendix~\ref{app:kappaExtraction}; we keep the notation \(\psi_\infty\) here to emphasize the bounded-trajectory interpretation. Within the controlled CMB window, the relevant modulus displacement is
\begin{equation}
\Delta\psi_{\rm CMB} \equiv \left|\psi_0(\varphi_*)-\psi_0(\varphi_{\rm ref})\right|,
\label{eq:deltaPsiCMB}
\end{equation}
where \(\varphi_{\rm ref}\) is a reference point in the same asymptotic window. This displacement can remain finite and parametrically small even when the canonical scalaron displacement is trans-Planckian. The end-of-inflation regime need not be accurately described by the same large-field series and should be followed with the full trajectory when precision near \(\varphi_{\rm end}\) is required. For \(\Delta\neq a\), the expansion is instead in powers of \(x^{\Delta/a}\), and the leading \(N\)-scaling of the tilt shift can change. The benchmark below, therefore, focuses on aligned, or effectively matched, trajectories whose first relevant threshold correction feeds the \(x\) and \(x^2\) plateau coefficients.

Along such a trajectory,
\begin{equation}
f_{\eff}(\varphi) = f\!\left(\psi_0(\varphi)\right) = f_\infty+f_1x+f_2x^2+\order(x^3),
\end{equation}
with
\begin{equation}
f_1=f_{,\psi}(\psi_\infty)A, \qquad f_2=f_{,\psi}(\psi_\infty)B+\frac12 f_{,\psi\psi}(\psi_\infty)A^2,
\end{equation}
where all derivatives are evaluated at the endpoint. The term \(f_\infty\GB\) is locally topological and irrelevant. The coefficients \(f_1,f_2,\ldots\) are not themselves observables; they are local threshold data that feed the matched single-clock deformation. Thus \(\kappa_G\) is a matched infrared parameter of the accepted trajectory, not a microscopic topological number.

The same expansion makes threshold freezing explicit:
\begin{equation}
\dot f_{\eff} = f_{,\psi}(\psi_0)\,\dot\psi_0 = f_{,\psi}(\psi_0) \frac{\dd\psi_0}{\dd\varphi}\dot\varphi \longrightarrow0 \qquad (\psi_0\to\psi_\infty).
\end{equation}
During the controlled slow approach, one should impose a slow variation of the threshold,
\begin{equation}
\left|\frac{\ddot f_{\eff}}{H\dot f_{\eff}}\right| = \order(\epsilon_H), \qquad \epsilon_H\equiv-\frac{\dot H}{H^2},
\end{equation}
while after the modulus settles both \(\dot f_{\eff}\) and \(\ddot f_{\eff}\) vanish. Thus, the threshold is allowed to vary during the CMB window, but only through controlled adiabatic settling of the heavy valley; it is not a freely rolling light field. Bounded-modulus mechanisms, including SHS-like constructions, motivate this finite-endpoint behavior \cite{Cagan:2025rbc}. More broadly, the ansatz is a compact EFT way to ask whether moduli can approach asymptotically controlled, supersymmetry-motivated endpoints without leaving the regime of four-dimensional effective-field-theory control.

\section{Dynamical background, attractor matching, and stability}
\label{sec:dynamicalanalysis}
This section derives the background equations for a dynamical scalar--Gauss--Bonnet threshold, explains precisely how the accepted two-field background is reduced to a matched single-clock plateau, and states the control conditions under which the resulting attractor formulae for \(\ns\), \(\rten\), and \(\alphas\) are meaningful.

\subsection{Background equations with a dynamical Gauss--Bonnet coupling}
\label{sec:bg}
For the spatially flat metric
\begin{equation}
\dd s^2=-\dd t^2+a(t)^2\dd\bm x^2,
\end{equation}
the Gauss--Bonnet density is
\begin{equation}
\GB = 24H^2(H^2+\dot H).
\label{eq:gbFLRW}
\end{equation}
For the background equations, it is convenient to write the coupling as \(\sigma_f f(\psi)\GB\), with \(\sigma_f=\pm1\). The convention used in \cref{eq:twofieldaction} corresponds to \(\sigma_f=-1\), and changing \(\sigma_f\) is equivalent to sending \(f\to -f\). We use a comma to denote partial differentiation with respect to the field indicated in the subscript, for example \(U_{,\varphi}\equiv\partial U/\partial\varphi\) and \(f_{,\psi}\equiv\partial f/\partial\psi\). With this normalization, the homogeneous equations are
\begin{align}
3H^2 &= \frac12\dot\varphi^2 +\frac12\dot\psi^2 +U -24\sigma_f H^3\dot f,
\label{eq:fullFriedmann}
\\ -2\dot H &= \dot\varphi^2 +\dot\psi^2 +8\sigma_f H^2\ddot f -8\sigma_f H^3\dot f +16\sigma_f H\dot H\dot f,
\label{eq:fullRay}
\\ \ddot\varphi+3H\dot\varphi+U_{,\varphi} &= 0,
\label{eq:fullPhi}
\\ \ddot\psi+3H\dot\psi+U_{,\psi} -\sigma_f f_{,\psi}\GB &= 0.
\label{eq:fullPsi}
\end{align}
Only the field-dependent part of the threshold enters these local equations, through \(\dot f\), \(\ddot f\), and \(f_{,\psi}\). The constant part of \(f\) multiplies the four-dimensional topological invariant and drops out of the local dynamics.

We use the Hubble slow-roll parameter
\begin{equation}
\epsilon_H\equiv-\frac{\dot H}{H^2}.
\end{equation}
In the slow-roll and weak-threshold regime,
\begin{equation}
\dot\varphi^2,\dot\psi^2\ll U, \qquad \epsilon_H\ll1, \qquad |H\dot f|\ll1, \qquad |\ddot f|\lesssim \order(\epsilon_H)|H\dot f|,
\label{eq:bgWeakRegime}
\end{equation}
the leading background equations reduce to
\begin{align}
3H^2 &\simeq U,
\label{eq:bgLeadingFriedmann}
\\
3H\dot\varphi+U_{,\varphi} &\simeq 0,
\label{eq:bgLeadingPhi}
\\
3H\dot\psi+U_{,\psi} -24\sigma_f H^4 f_{,\psi} &\simeq 0.
\label{eq:bgLeadingPsi}
\end{align}
If the threshold-active field is heavy, it is not an independent light slow-roll field. Instead, it adiabatically tracks the instantaneous Gauss--Bonnet-shifted minimum,
\begin{equation}
U_{,\psi}(\varphi,\psi_0) - 24\sigma_f H^4(\varphi,\psi_0)\, f_{,\psi}(\psi_0) = 0.
\label{eq:minimum}
\end{equation}
The relevant local curvature of this shifted minimum is
\begin{equation}
m_{\psi,\eff}^2 \equiv U_{,\psi\psi} -24\sigma_f H^4 f_{,\psi\psi},
\label{eq:meffPsiMain}
\end{equation}
evaluated on the tracked solution \(\psi=\psi_0(\varphi)\). The single-clock limit requires this mass hierarchy together with small kinetic mixing and small turn rate, as summarized in
\cref{sec:stability}.

The accepted adiabatic background is then described by the trajectory
\begin{equation}
\psi=\psi_0(\varphi),
\label{eq:adiabaticTrajectory}
\end{equation}
and by the two functions
\begin{equation}
U_{\eff}(\varphi) \equiv U\!\left(\varphi,\psi_0(\varphi)\right), \qquad f_{\eff}(\varphi) \equiv f\!\left(\psi_0(\varphi)\right).
\label{eq:effectiveFunctionsMain}
\end{equation}
More precisely, in the weak-threshold regime, the accepted background satisfies
\begin{equation}
3H_{\rm ad}^2(\varphi) = U_{\eff}(\varphi) \left[ 1+\order\!\left(\epsilon_H,H\dot f_{\eff}\right) \right],
\label{eq:UeffBackgroundMatching}
\end{equation}
and the scalaron drift can be written in the matched single-clock form
\begin{equation}
3H_{\rm ad}\dot\varphi + \frac{\dd U_{\eff}}{\dd\varphi} = \Delta_{\rm turn} + \Delta_{\rm GB},
\label{eq:UeffDriftMatching}
\end{equation}
where \(\Delta_{\rm turn}\) denotes corrections induced by the bending of the two-field trajectory and \(\Delta_{\rm GB}\) denotes direct scalar--Gauss--Bonnet corrections not captured by the plateau-shape matching. In the regime used below, these terms are subleading,
\begin{equation}
\frac{|\Delta_{\rm turn}|+|\Delta_{\rm GB}|} {|3H_{\rm ad}\dot\varphi|} \ll1.
\label{eq:driftCorrectionControl}
\end{equation}
Thus \(U_{\eff}\) is an operational single-clock background function: it is the plateau function that matches \(3H^2\) and the leading adiabatic drift of the solved background.

It is important to emphasize what is, and is not, meant by \(U_{\eff}\). The Gauss--Bonnet term has not been converted into an ordinary scalar potential by substituting \(\GB=\GB_{\rm FLRW}\). Rather, the scalar--Gauss--Bonnet force shifts the instantaneous minimum of the heavy threshold field. Solving this shifted minimum gives the adiabatic trajectory \(\psi_0(\varphi)\), and \(U_{\eff}(\varphi)=U(\varphi,\psi_0(\varphi))\) is the matched single-clock background function along that trajectory. This construction is valid only when the direct contributions controlled by \(H\dot f_{\eff}\), \(\ddot f_{\eff}\), and entropic transfer are perturbative. In this regime, the leading observable effect can be encoded in the invariant plateau coefficient \(A_G\). Outside this regime, the spectra must be computed from the full scalar--Gauss--Bonnet quadratic action.

The practical adiabatic-control conditions used in the matching are
\begin{equation}
\frac{m_{\psi,\eff}^2}{H^2}\gg1, \qquad \left|\frac{\dd\psi_0}{\dd\varphi}\right|\ll1, \qquad |H\dot f_{\eff}|\ll1.
\label{eq:practicalAdiabaticControl}
\end{equation}
Under these conditions, the accepted background can be summarized by the invariant plateau coefficient introduced next.

\subsection{Invariant attractor expansion and observable derivation}
\label{sec:derivation}
The calculable correction should not be described by forcing an off-shell \(f(R,\GB)\) Einstein-frame map. Instead, the accepted single-clock background is matched directly to its large-field plateau expansion. Define
\begin{equation}
x=e^{-a\varphi}, \qquad a=\sqrt{\frac23}.
\label{eq:xDefinitionMain}
\end{equation}
The large-field endpoint corresponds to \(x\to0\). In reduced Planck units, the plateau height is defined by
\begin{equation}
\Lambda^4 \equiv \lim_{\varphi\to\infty}U_{\eff}(\varphi), \qquad H_0^2 \equiv \frac{\Lambda^4}{3}.
\label{eq:LambdaDefinitionMain}
\end{equation}
Here \(H_0\) is the asymptotic quasi-de Sitter Hubble scale of the matched single-clock background. In the pure Starobinsky limit,
\begin{equation}
U_S(\varphi) = \frac{3M^2}{4} \left(1-e^{-a\varphi}\right)^2, \qquad \Lambda^4=\frac{3M^2}{4}, \qquad H_0^2=\frac{M^2}{4}.
\label{eq:StarobinskyLambdaH0}
\end{equation}
Suppose the matched plateau has the asymptotic form
\begin{equation}
U_{\eff}(\varphi) = \Lambda^4 \left[ 1+u_1x+u_2x^2+\order(x^3) \right], \qquad u_1<0.
\label{eq:generalMatchedPlateau}
\end{equation}
A constant shift of the canonical scalaron rescales \(x\), so \(u_2\) alone is not invariant. The leading falloff is fixed by defining
\begin{equation}
\tilde x=-\frac{u_1}{2}x.
\label{eq:normalizedXMain}
\end{equation}
Then the same plateau can be written as
\begin{equation}
U_{\eff}(\varphi) = \Lambda^4 \left[ 1-2\tilde x+A_G\tilde x^2+\order(\tilde x^3) \right], \qquad A_G=\frac{4u_2}{u_1^2}.
\label{eq:plateauA}
\end{equation}
Thus \(A_G\) is invariant under constant scalaron shifts. For the pure Starobinsky potential, \(u_1=-2\), \(u_2=1\), and \(A_G=1\). We parameterize the departure from Starobinsky by
\begin{equation}
A_G=1+\frac{2\kappa_G}{3}.
\label{eq:kappaDefinitionMain}
\end{equation}
The coefficients \(u_1,u_2\), and hence \(\kappa_G\), are matched to the infrared data of the accepted single-clock trajectory. They can receive contributions from bounded threshold motion, heavy-field elimination, and perturbative scalar--Gauss--Bonnet backreaction. They are not topological numbers, and they should not be interpreted as coefficients of an off-shell potential obtained by substituting the background value of \(\GB\).

If the deformation relative to the Starobinsky coefficients is written as
\begin{equation}
u_1=-2+\delta u_1,
\qquad
u_2=1+\delta u_2,
\label{eq:deltaUiMain}
\end{equation}
then, to first order,
\begin{equation}
A_G-1 \simeq \delta u_1+\delta u_2, \qquad \kappa_G \simeq \frac32(\delta u_1+\delta u_2).
\label{eq:linearizedKappaMain}
\end{equation}
This form makes explicit that threshold-induced shifts in both the leading and subleading plateau coefficients contribute to the invariant deformation.

In the weak-threshold regime, this also gives the direct bridge to compactification data. Let \(\lambda_{\rm GB}\) denote the microscopic normalization of the field-dependent Gauss--Bonnet threshold, as in \cref{eq:lambdaGeneral}. After solving the shifted heavy-field trajectory and matching the resulting single-clock background, the plateau coefficients may be expanded as
\begin{equation}
u_i = u_i^{(0)} + \lambda_{\rm GB}\,\widehat u_i + \order(\lambda_{\rm GB}^2), \qquad u_1^{(0)}=-2, \qquad u_2^{(0)}=1,
\label{eq:uiThresholdResponse}
\end{equation}
where the response coefficients \(\widehat u_i\) are not universal numbers. They encode the local threshold derivatives, the stabilization sector, the trajectory orientation in moduli space, and the single-clock matching prescription. Substituting
\cref{eq:uiThresholdResponse} into \cref{eq:linearizedKappaMain} gives
\begin{equation}
\kappa_G \simeq \frac32\,\lambda_{\rm GB} \left( \widehat u_1+\widehat u_2 \right) + \order(\lambda_{\rm GB}^2).
\label{eq:kappaFromThresholdResponse}
\end{equation}
Using the schematic compactification prior \(\lambda_{\rm GB}\sim c_{\rm GB}\mathcal P(g_s,\calV_s,\alpha')\mathcal T(K_6)\), this becomes
\begin{equation}
\kappa_G \simeq \mathcal C_{\rm eff}\,\mathcal T(K_6), \qquad \mathcal C_{\rm eff} \equiv \frac32\, c_{\rm GB}\mathcal P(g_s,\calV_s,\alpha') \left( \widehat u_1+\widehat u_2 \right),
\label{eq:ceffFromPlateauResponse}
\end{equation}
up to higher-order threshold corrections and convention-dependent normalizations. Thus \(\mathcal C_{\rm eff}\) is simply the compact notation for the response of the matched plateau coefficients to the microscopic threshold data. It is not an additional universal parameter, and it cannot be fixed by topology alone.

In what follows \(x\) denotes the normalized variable \(\tilde x\), and the matched potential is written as
\begin{equation}
U_{\eff}(\varphi) = \Lambda^4 \left[ 1-2x+A_Gx^2+\order(x^3) \right].
\label{eq:normalizedPlateauMain}
\end{equation}
For admissible threshold trajectories of the form \cref{eq:admissiblePsi}, the local threshold data \(f_1,f_2,\ldots\) enter the observable calculation only through such matched coefficients. Appendix~\ref{app:kappaExtraction} summarizes how \(\kappa_G\) is extracted from a solved adiabatic background, either analytically or in a numerical scan.

The formulas below assume
\begin{equation}
|H\dot f_{\eff}|\ll1, \qquad \left|\frac{\ddot f_{\eff}}{H\dot f_{\eff}}\right|=\order(\epsilon_H), \qquad c_s^2\simeq c_T^2\simeq1,
\label{eq:weakGBPerturbativeMain}
\end{equation}
so that direct Gauss--Bonnet corrections to the quadratic perturbation action are subleading relative to the plateau-shape deformation. In this controlled regime, one may use the usual single-field potential slow-roll expressions as an attractor diagnostic,
\begin{equation}
\epsilon_V = \frac12 \left( \frac{U_{\eff,\varphi}}{U_{\eff}} \right)^2, \qquad \eta_V =\frac{U_{\eff,\varphi\varphi}}{U_{\eff}}, \qquad \ns\simeq1-6\epsilon_V+2\eta_V, \qquad \rten\simeq16\epsilon_V.
\label{eq:potentialSlowRollDefinitionsMain}
\end{equation}
For \cref{eq:normalizedPlateauMain}, one finds
\begin{align}
\epsilon_V &= \frac43 x^2 + \frac83(2-A_G)x^3 +\order(x^4),
\label{eq:epsilonExpansionMain}
\\
\eta_V
&= -\frac43 x + \frac83(A_G-1)x^2 +\order(x^3).
\label{eq:etaExpansionMain}
\end{align}
The leading e-fold relation is unchanged at the order needed for the \(A_G\)-dependent shift,
\begin{equation}
x_* = \frac{3}{4N} + \order\!\left(\frac{\ln N}{N^2}\right),
\label{eq:xstarMain}
\end{equation}
where \(N\) counts e-folds before the end of inflation. Substituting \cref{eq:xstarMain} into \cref{eq:potentialSlowRollDefinitionsMain} and subtracting the Starobinsky result at the same \(N\), one obtains
\begin{equation}
\Delta\ns \equiv \ns-\ns^{(0)} = \frac{3(A_G-1)}{N^2} + \order\!\left(\frac{\ln N}{N^3}\right) = \frac{2\kappa_G}{N^2} +
\order\!\left(\frac{\ln N}{N^3}\right).
\label{eq:dnsmasterMain}
\end{equation}
Similarly,
\begin{equation}
\rten = \rten^{(0)}(N) \left[ 1-\frac{3(A_G-1)}{2N} +\order(N^{-2}) \right] = \rten^{(0)}(N) \left[ 1-\frac{\kappa_G}{N} \right] +\order(N^{-4}),
\label{eq:rmasterMain}
\end{equation}
with
\begin{equation}
\rten^{(0)}(N) = \frac{12}{N^2} + \order\!\left(\frac{\ln N}{N^3}\right).
\label{eq:r0Main}
\end{equation}
Since \(N\) counts e-folds before the end of inflation, \(\alphas\equiv\dd\ns/\dd\ln k\simeq-\dd\ns/\dd N\). For
\(\ns(N)\simeq1-2/N+2\kappa_G(N)/N^2\), this gives
\begin{equation}
\alphas = -\frac{2}{N^2} + \frac{4\kappa_G}{N^3} - \frac{2}{N^2}\frac{\dd\kappa_G}{\dd N}.
\label{eq:alphasGeneral}
\end{equation}
The benchmark applications assume that \(\kappa_G\) is approximately constant across the CMB window, so that
\begin{equation}
\alphas \simeq -\frac{2}{N^2} + \frac{4\kappa_G}{N^3}.
\label{eq:alphasConstantKappaMain}
\end{equation}
If \(\kappa_G\) varies appreciably, the last term in \cref{eq:alphasGeneral} must be retained.

The central analytic output is therefore
\begin{equation}
\boxed{ \Delta\ns \simeq \frac{2\kappa_G}{N^2}, \qquad \rten \simeq \frac{12}{N^2} \left( 1-\frac{\kappa_G}{N} \right), \qquad \alphas \simeq -\frac{2}{N^2} + \frac{4\kappa_G}{N^3}}
\label{eq:centralResults}
\end{equation}
within the controlled single-clock, weak-threshold regime. These expressions depend on the matched plateau expansion and not on any constant \(\xi\GB\) deformation. Outside the regime specified in \cref{eq:weakGBPerturbativeMain}, one must compute \(\ns\), \(\rten\), and \(\alphas\) from the full scalar--Gauss--Bonnet perturbation action rather than from the potential slow-roll expressions.

The same formulae can be inverted. For a target scalar tilt \(\ns^{\rm tar}\) at fixed \(N\),
\begin{equation}
\kappa_G(N,\ns^{\rm tar}) = \frac{N^2}{2} \left( \ns^{\rm tar}-1+\frac2N \right),
\label{eq:kappaTarget}
\end{equation}
and the associated tensor prediction is
\begin{equation}
\rten(N,\ns^{\rm tar}) = \frac{12}{N^2} \left[ 1-\frac{N}{2} \left( \ns^{\rm tar}-1+\frac2N \right)\right].
\label{eq:rTarget}
\end{equation}
For constant \(\kappa_G\), the corresponding running is
\begin{equation}
\alphas(N,\ns^{\rm tar}) = -\frac{2}{N^2} + \frac{2}{N} \left( \ns^{\rm tar}-1+\frac2N \right) = \frac{2(\ns^{\rm tar}-1)}{N} + \frac{2}{N^2}.
\label{eq:alphasTarget}
\end{equation}
Thus CMB data select a target range for \(\kappa_G\), while explicit threshold trajectories determine whether that range can be reached in a controlled compactification.

\subsection{Validity conditions for the matched benchmark}
\label{sec:stability}
The benchmark formulae are meaningful only inside a controlled single-clock regime. The scalar and tensor quadratic actions must first be free of ghost and gradient instabilities,
\begin{equation}
Q_T>0, \qquad c_T^2>0, \qquad Q_s>0, \qquad c_s^2>0.
\label{eq:stability}
\end{equation}
Direct scalar--Gauss--Bonnet corrections must also be perturbative. We measure them by
\begin{equation}
\delta_{\rm GB} \equiv H\dot f_{\eff},
\label{eq:deltaGBMain}
\end{equation}
and require
\begin{equation}
|\delta_{\rm GB}|\ll1, \qquad \left| \frac{\dot\delta_{\rm GB}}{H\delta_{\rm GB}} \right| =\order(\epsilon_H).
\label{eq:smallDeltaGB}
\end{equation}
The first condition keeps the direct scalar--Gauss--Bonnet modification of the quadratic action subleading. The second condition excludes sharp threshold features in the smooth-attractor benchmark.

The orthogonal fluctuation must decouple from the observed adiabatic mode \cite{Gordon:2000hv,Achucarro:2010da}. Thus, the heavy-modulus condition is an entropic decoupling condition, not merely a statement about an isolated mass eigenvalue. If \(M_{\rm ent}\) is the entropic mass and \(\Omega_{\rm turn}\) is the covariant turn rate of the adiabatic tangent, a sufficient single-clock hierarchy is
\begin{equation}
\frac{M_{\rm ent}^2}{H^2}\gg1, \qquad \frac{\Omega_{\rm turn}^2}{M_{\rm ent}^2}\ll1.
\label{eq:entropicHierarchyMain}
\end{equation}
In the threshold-matching regime, this is implemented by
\begin{equation}
\frac{m_{\psi,\eff}^2}{H^2}\gg1, \qquad \left|\frac{\dd\psi_0}{\dd\varphi}\right|\ll1, \qquad |H\dot f_{\eff}|\ll1.
\label{eq:practicalSingleClockMain}
\end{equation}
Together with the perturbation-health conditions in \cref{eq:stability}, this hierarchy permits the use of a single invariant plateau coefficient \(A_G\) rather than a two-field transfer-function calculation.

Finally, the compactification and attractor expansions must remain below their cutoffs:
\begin{equation}
\frac{H_*^2}{M_{\rm KK}^2}\ll1, \qquad \frac{H_*^2}{M_s^2}\ll1, \qquad \frac{H_*^2}{m_{\psi,\eff}^2}\ll1, \qquad \frac{|\kappa_G|}{N}\ll1.
\label{eq:eftHierarchy}
\end{equation}
The last condition is the perturbative attractor requirement: when \(|\kappa_G|/N\) is no longer small, the neglected \(\order(x^3)\) coefficients, possible \(N\)-dependence of \(\kappa_G\), and direct scalar--Gauss--Bonnet perturbation effects must be checked explicitly. A benchmark point is physically meaningful only if it passes the health, single-clock, and cutoff cuts above. Outside this regime, the matched formulae in \cref{eq:centralResults} are diagnostics only, and the spectra must be computed from the full scalar--Gauss--Bonnet perturbation system.

A viable cosmological completion must also freeze the threshold after reheating,
\begin{equation}
\dot f\rightarrow0, \qquad \ddot f\rightarrow0,
\label{eq:lateTimeFreeze}
\end{equation}
so that the late-time tensor speed returns to its general-relativistic value.

\section{Phenomenology, reheating, and scan protocol}
\label{sec:phenomenology}
This section translates the invariant attractor coefficient into benchmark observables, then uses representative compactification data only as a target diagnostic for future explicit models.

\subsection{Numerical benchmarks against current data}
\label{sec:benchmarks}
The observational anchors used below are Planck 2018, BICEP/Keck BK18, and recent combined CMB+BAO/SPT/ACT/DESI constraints
\cite{Planck:2018vyg,Planck:2018jri,BICEP:2021xfz,Balkenhol:2025wms}. Planck provides the Starobinsky-compatible baseline, BK18 gives the tensor-safety constraint, and recent ACT/SPT-informed combinations motivate the higher-\(\ns\) target direction
\cite{AtacamaCosmologyTelescope:2025blo,Balkenhol:2025wms}. We use these data only as benchmark anchors; no new combined likelihood analysis is performed. A related recent discussion of plateau and attractor models in light of the higher-\(\ns\) direction appears in
\cite{Kallosh:2025ijd}. Current constraints on the running are much weaker than those on the tilt, and the values predicted below remain of order \(10^{-4}\)--\(10^{-3}\) in the controlled benchmark regime.

\begin{figure}[htb]
\centering
\includegraphics[width=1.0\textwidth]{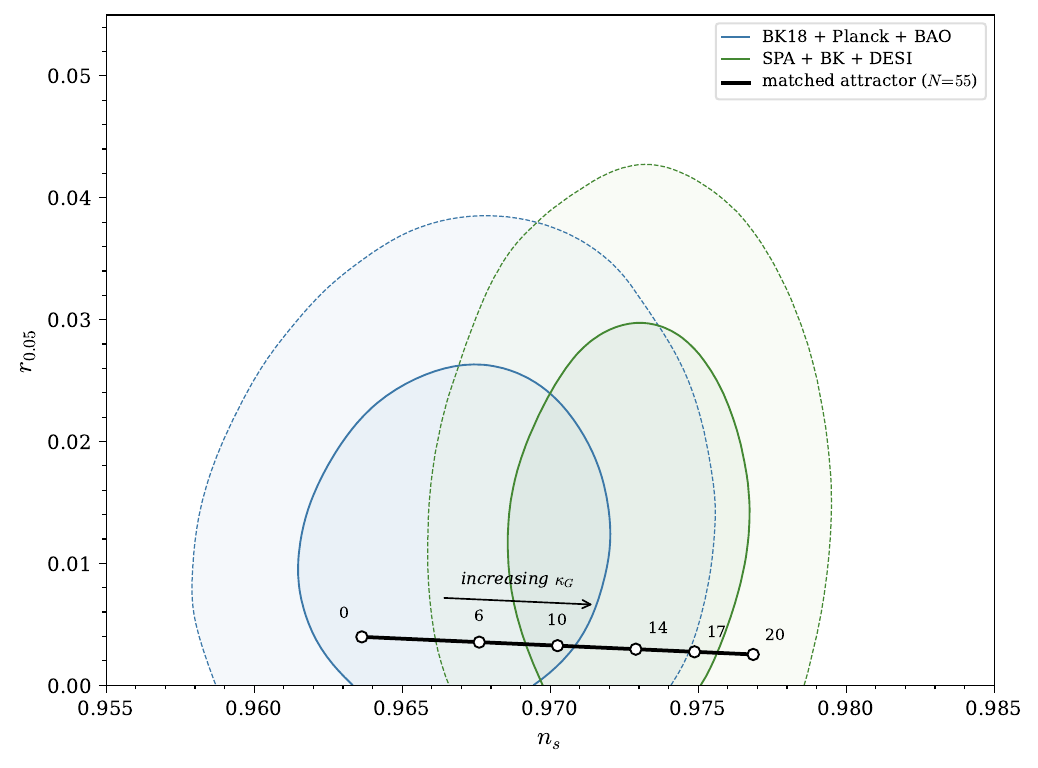}
\caption{Matched-attractor trajectory in the \((\ns,\rten)\) plane at \(N=55\), with \(\kappa_G\) varied continuously from \(0\) to \(20\). Increasing \(\kappa_G\) raises \(\ns\) and lowers \(\rten\). The labels along the trajectory denote selected values of \(\kappa_G\). The contours show the 68\% and 95\% credible regions for the public Planck/BK18 \cite{Planck:2018vyg,Planck:2018jri,BICEP:2021xfz} and the recent combined SPT+Planck+ACT+BICEP/Keck+DESI data products \cite{Balkenhol:2025wms}.}
\label{fig:nsrKappaContours}
\end{figure}

For numerical illustration, we use the leading attractor formulae
\begin{equation}
\ns \simeq 1-\frac2N+\frac{2\kappa_G}{N^2}, \qquad \rten\simeq\frac{12}{N^2}\left(1-\frac{\kappa_G}{N}\right), \qquad \alphas\simeq-\frac{2}{N^2}+\frac{4\kappa_G}{N^3}.
\label{eq:numericformula}
\end{equation}
These expressions are valid only in the controlled regime described in \cref{sec:stability}. The inverse map in \cref{eq:kappaTarget,eq:rTarget,eq:alphasTarget} gives the target-space values in \cref{tab:targetKappa}. The continuous forward trajectory at \(N=55\) is shown in \cref{fig:nsrKappaContours}. Increasing \(\kappa_G\) raises \(\ns\), lowers \(\rten\), and makes \(\alphas\) less negative in the constant-\(\kappa_G\) approximation.
\begin{table}[htb]
\centering
\caption{Inverse attractor matching at \(N=55\). Each row asks what effective plateau deformation is needed to hit a chosen scalar-tilt target, then reports the associated tensor prediction and constant-\(\kappa_G\) running.}
\begin{tabular}{ccccc}
\toprule
Target \(\ns^{\rm tar}\) & Motivation & Required \(\kappa_G\) & \(\rten\) & \(\alphas\)\\
\midrule
0.9682 & recent CMB & 6.90 & 0.00347 & \(-4.95\times10^{-4}\)\\
0.9728 & recent CMB+BAO & 13.86 & 0.00297 & \(-3.28\times10^{-4}\)\\
0.9740 & ACT-like & 15.68 & 0.00284 & \(-2.84\times10^{-4}\)\\
0.9748 & high edge & 16.89 & 0.00275 & \(-2.55\times10^{-4}\)\\
\bottomrule
\end{tabular}
\label{tab:targetKappa}
\end{table}
The figure shows the continuous version of the benchmark trajectory. Positive \(\kappa_G\) moves the Starobinsky point toward larger \(\ns\) and smaller \(\rten\), while \(\alphas\) remains of order \(10^{-4}\)--\(10^{-3}\) across the plotted interval. The higher-\(\ns\) target direction is represented here by the recent combined SPT+Planck+ACT+BICEP/Keck+DESI contour, abbreviated as SPA+BK+DESI in the data release \cite{Balkenhol:2025wms}, while the ACT DR6 result provides additional motivation for this direction \cite{AtacamaCosmologyTelescope:2025blo}.

Changing \(N\) within the usual \(50\)--\(60\) reheating window shifts all entries in the standard way, so the benchmark trajectory should not be interpreted without an \(N_*\) prior. The high-\(\kappa_G\) end of the curve requires stronger control in the sense that
\(|\kappa_G|/N\) is no longer parametrically very small, so the neglected \(\order(x^3)\) plateau coefficients, possible \(N\)-dependence of \(\kappa_G\), and direct scalar--Gauss--Bonnet corrections to the quadratic action must be checked explicitly.

\subsection{Semi-microscopic target diagnostic}
\label{subsec:workedTopologyBenchmark}
The compactification weights in \cref{tab:cyprior} can be connected to the observable benchmark through the response of the matched plateau coefficients. In the weak-threshold expansion of \cref{eq:uiThresholdResponse,eq:kappaFromThresholdResponse}, the microscopic threshold normalization shifts \(u_1\) and \(u_2\), and therefore shifts the invariant combination \(A_G=4u_2/u_1^2\). If the leading topology dependence of the threshold normalization is packaged as \(\lambda_{\rm GB}\sim c_{\rm GB}\mathcal P\,\mathcal T(K_6)\), the result can be written as
\begin{equation}
\kappa_G\simeq \mathcal C_{\rm eff}\mathcal T(K_6),
\label{eq:kappaCeffT}
\end{equation}
with \(\mathcal C_{\rm eff}\) defined by the plateau-response relation \cref{eq:ceffFromPlateauResponse}. This relation should be read as a target diagnostic, not as a topological prediction. The topological factor \(\mathcal T(K_6)\) supplies only a discrete prior. The coefficient \(\mathcal C_{\rm eff}\) packages the remaining continuous and dynamical data: the normalization of the four-dimensional threshold, local derivatives of the string threshold function, the stabilized modulus mass matrix, the orientation of the accepted trajectory in moduli space, possible flux and warping dependence, and the single-clock matching from the solved background to \(A_G\) and \(\kappa_G\).

Equivalently, once a compactification and stabilized trajectory have been specified, the nonlinear matched value is obtained from the invariant plateau combination
\begin{equation}
\kappa_G = \frac32 \left[ \frac{4u_2}{u_1^2} - 1 \right],
\label{eq:kappaExactFromUi}
\end{equation}
where \(u_1\) and \(u_2\) are extracted from the solved single-clock background in the CMB asymptotic window. The linear relation
\(\kappa_G\simeq\mathcal C_{\rm eff}\mathcal T(K_6)\) is the first-order, weak-threshold form of this matching prescription.

For the \(X_{24}(1,1,2,8,12)\) benchmark, \(\EulerChar(K_6)=-480\), so that
\begin{equation}
\mathcal T(K_6)=\frac{\EulerChar(K_6)}{24}=-20.
\end{equation}
Choosing the target effective response \(\mathcal C_{\rm eff}=-0.70\) gives
\begin{equation}
\kappa_G \simeq \mathcal C_{\rm eff}\mathcal T(K_6) = (-0.70)(-20) = 14.
\end{equation}
The full interval \(\kappa_G\simeq7\)--\(17\) maps to \(\mathcal C_{\rm eff}\simeq-0.35\)--\(-0.85\) for this topology. Values of
\(\mathcal C_{\rm eff}\) of order unity are not excluded at the level of effective threshold matching, but they are not guaranteed by topology alone. In particular, fluxes, warping, bundle data, and the stabilization sector can change both the magnitude and the sign of the continuous response. Therefore the \(X_{24}\) example should be read only as a target diagnostic: \(\mathcal C_{\rm eff}\) must be computed in an explicit stabilized compactification and then checked against the single-clock, weak-\(\delta_{\rm GB}\), stability, and cutoff conditions of
\cref{sec:stability}.

At \(N=55\), the target value \(\kappa_G=14\) gives
\begin{align}
\ns &\simeq 1-\frac{2}{55}+\frac{2(14)}{55^2} = 0.9729, \\ \rten
&\simeq \frac{12}{55^2} \left( 1-\frac{14}{55}\right)=2.96\times10^{-3},
\\ \alphas
&\simeq -\frac{2}{55^2} + \frac{4(14)}{55^3} =-3.24\times10^{-4}.
\end{align}
Thus, a representative CY topological weight, combined with a target effective response, reproduces the central benchmark value at the level of the matched-attractor parametrization. A controlled compactification realization remains to be demonstrated.

\subsection{Reheating uncertainty and numerical scan protocol}
The pivot-to-\(N_*\) mapping depends on reheating, with the usual Starobinsky-like range \(N_*\simeq50\)--\(60\). In the present model, reheating must also lead to a vacuum in which the threshold-active modulus is frozen, as in \cref{eq:lateTimeFreeze}. The benchmark table and trajectory should therefore be read together with the usual \(N_*\) uncertainty.

A full numerical construction should:
\begin{enumerate}
\item impose the compactification hierarchy \(H_*^2/M_s^2\ll1\) and \(H_*^2/M_{\rm KK}^2\ll1\), or equivalently \cref{eq:stringCutoffBound,eq:kkCutoffBound};
\item choose the topological weight \(\mathcal T(K_6)\), the threshold-shape function \(\mathcal F(\psi_i)\), and the stabilization sector, then determine the response coefficient relating \(\lambda_{\rm GB}\) to the matched \(\kappa_G\);
\item solve the full background equations \cref{eq:fullFriedmann,eq:fullRay,eq:fullPhi,eq:fullPsi} and map the pivot scale to \(N_*\),
including reheating;
\item construct the adiabatic/entropic decomposition, track \(M_{\rm ent}^2/H^2\), \(\Omega_{\rm turn}^2/M_{\rm ent}^2\), and \(|\dd\psi_0/\dd\varphi|\), and require a single-clock valley before the CMB window exits;
\item compute \(Q_T,c_T^2,Q_s,c_s^2\) from the scalar--Gauss--Bonnet quadratic action and impose ghost and gradient stability;
\item extract \(\As,\ns,\rten,\alphas\) from the properly normalized spectra, using \cref{eq:numericformula} only as a weak-coupling attractor diagnostic;
\item verify bounded threshold motion, including the displacement \(\Delta\psi_{\rm CMB}\) in \cref{eq:deltaPsiCMB}, approach to a finite endpoint, late-time freezing \(\dot f,\ddot f\to0\), and the absence of sharp threshold features in the smooth-plateau benchmark;
\item match the accepted trajectory to \(A_G\), \(\kappa_G\), and \(\alphas\).
\end{enumerate}
The central theoretical test is whether the interval selected in \cref{tab:targetKappa}, \(\kappa_G\simeq7\)--\(17\), is dynamically attainable while maintaining \cref{eq:smallDeltaGB,eq:eftHierarchy}. Localized threshold features may be interesting for small-scale power enhancement, but they lie outside the smooth-plateau benchmark considered here.

\section{Conclusions and outlook}
\label{sec:discussionconclusions}
We have developed a controlled, effective dynamical Gauss--Bonnet deformation of Starobinsky inflation in which the correction is sourced by a stabilized threshold field and then matched to a single-clock plateau. The construction combines three ingredients: a frame-safe Starobinsky scalaron baseline, a bounded stabilized threshold trajectory, and a scalar-dependent Gauss--Bonnet coupling. A constant four-dimensional Gauss--Bonnet term remains topological and cannot shift local inflationary observables; the relevant physical input is the variation of the threshold function along the accepted trajectory,
\begin{equation}
f_{,\varphi}^{\rm eff} = f_{,\psi} \frac{\dd\psi_0}{\dd\varphi}.
\end{equation}
Equivalently, the Horndeski/Galileon language is used only as the induced low-energy representation of this scalar--Gauss--Bonnet threshold after heavy-field stabilization, not as an independent phenomenological sector. This is also what separates the present construction from bare regularized four-dimensional Einstein--Gauss--Bonnet prescriptions: local four-dimensional dynamics is generated here by the field dependence of \(f(\psi)\), rather than by assigning local meaning to a constant rescaled Gauss--Bonnet coupling. Calabi--Yau topology can therefore enter only as a prior on the threshold function. It does not directly predict \(\ns\), \(\rten\), or \(\alphas\). The parameter \(\kappa_G\) is instead a matched infrared plateau deformation of the accepted single-clock trajectory.

The resulting large-field background is summarized by
\begin{equation}
U_{\eff}(\varphi) = \Lambda^4 \left[ 1 -2e^{-\sqrt{2/3}\varphi} + \left( 1+\frac{2\kappa_G}{3} \right) e^{-2\sqrt{2/3}\varphi} +\cdots \right].
\end{equation}
The leading attractor-level predictions are
\begin{equation}
\Delta\ns = \frac{2\kappa_G}{N^2}, \qquad \rten = \frac{12}{N^2} \left( 1-\frac{\kappa_G}{N} \right), \qquad \alphas \simeq -\frac{2}{N^2} + \frac{4\kappa_G}{N^3},
\end{equation}
up to higher-order attractor, reheating, and direct scalar--Gauss--Bonnet perturbation effects. For \(N\simeq55\), the benchmark interval
\begin{equation}
\kappa_G\simeq7\text{--}17
\end{equation}
raises the scalar tilt while keeping \(\rten\simeq0.003\) or below and \(|\alphas|\lesssim10^{-3}\). This is a benchmark trajectory, not a likelihood analysis.

The interval is useful only if it can be attained while maintaining the full control regime:
\begin{equation}
|H\dot f_{\eff}|\ll1, \qquad \left|\frac{\ddot f_{\eff}}{H\dot f_{\eff}}\right|=\order(\epsilon_H), \qquad \frac{m_{\psi,\eff}^2}{H^2}\gg1, \qquad \frac{H^2}{M_s^2}\ll1, \qquad \frac{H^2}{M_{\rm KK}^2}\ll1,
\end{equation}
together with healthy scalar and tensor quadratic actions. The \(X_{24}(1,1,2,8,12)\) example gives a quantitative compactification target, not a completed compactification prediction: \(\mathcal C_{\rm eff}\) must be computed from the threshold function, stabilization data, and the accepted single-clock trajectory. The natural next step is to construct explicit stabilized compactifications, or controlled threshold EFTs with the same microscopic data, that realize the target range \(\kappa_G\simeq7\text{--}17\), verify the single-clock and weak-coupling conditions, and confront the resulting spectra with full CMB likelihoods.

\section*{Acknowledgments}
The author thanks Mehmet Ozkan for earlier discussions on Galileon inflation that motivated the original question, and Cemal Berfu Senisik for useful discussions on Gauss--Bonnet couplings. The contribution of O.G. was supported in part by the Istanbul Technical University Research Fund under grant number 2025-47239.

\appendix

\section{Starobinsky baseline formulae}
\label{app:starobinskyBaseline}

This appendix collects the standard Starobinsky formulae used in \cref{sec:starobinsky}. The \(R^2\) term may be linearized by
\begin{equation}
\frac12\left(R+\frac{R^2}{6M^2}\right) \quad\longleftrightarrow\quad \frac12\left[ \Phi R-\frac{3M^2}{2}(\Phi-1)^2 \right],
\end{equation}
whose algebraic equation gives \(\Phi=1+R/(3M^2)\). With \(\Phi=e^{\sqrt{2/3}\varphi}\), the Einstein-frame potential is \cref{eq:staropot}. Defining
\begin{equation}
x=e^{-\sqrt{2/3}\varphi},
\end{equation}
one finds
\begin{equation}
\epsilon_V = \frac43\frac{x^2}{(1-x)^2}, \qquad x_{\rm end} = \frac{\sqrt3}{2+\sqrt3}.
\end{equation}
The exact e-fold relation is
\begin{equation}
N = \frac34 \left( \frac{1}{x_*}+\ln x_* \right) - \frac34 \left(\frac{1}{x_{\rm end}}+\ln x_{\rm end}\right),
\end{equation}
and therefore
\begin{equation}
x_*= \frac{3}{4N} + \order\!\left(\frac{\ln N}{N^2}\right).
\end{equation}
This gives the standard large-\(N\) predictions
\begin{equation}
\ns^{(0)} = 1-\frac2N + \order\!\left(\frac{\ln N}{N^2}\right), \qquad \rten^{(0)} = \frac{12}{N^2} + \order\!\left(\frac{\ln N}{N^3}\right), \qquad \alphas^{(0)} = -\frac2{N^2} + \order\!\left(\frac{\ln N}{N^3}\right).
\end{equation}
The scalar amplitude fixes the Starobinsky mass scale:
\begin{equation}
\As = \frac{U_*}{24\pi^2\epsilon_*} \simeq \frac{M^2N^2}{24\pi^2}.
\end{equation}
For \(\As\simeq2.1\times10^{-9}\) and \(N=55\), this gives
\begin{equation}
M\simeq1.3\times10^{-5}.
\end{equation}
Since \(U_S=(3M^2/4)(1-x)^2\), the plateau height is \(\Lambda^4=3M^2/4\), and the associated quasi-de Sitter scale is
\begin{equation}
H_0^2=\frac{\Lambda^4}{3} = \frac{M^2}{4} \simeq \frac{6\pi^2\As}{N^2}.
\end{equation}
For the same numerical values, \(H_0^2\simeq4.1\times10^{-11}\) and \(H_0\simeq6.4\times10^{-6}\) in reduced Planck units.

\section{Higher-curvature comparison}
\label{app:higherCurvatureComparison}
This appendix records the simple \(f(R)\) comparison quoted in \cref{sec:starobinsky}. For
\begin{equation}
f(R)=R+\alpha R^2+\beta R^3,
\end{equation}
one has
\begin{equation}
f_{,R} = 1+2\alpha R+3\beta R^2, \qquad f_{,R}=e^{\sqrt{2/3}\varphi}.
\end{equation}
The Einstein-frame potential is
\begin{equation}
U(\varphi)=\frac{R f_{,R}-f}{2f_{,R}^2}.
\label{eq:fRpotential}
\end{equation}
Solving perturbatively in \(\beta\), with \(x=e^{-\sqrt{2/3}\varphi}=f_{,R}^{-1}\), gives
\begin{equation}
R= \frac{1-x}{2\alpha x} - \frac{3\beta}{8\alpha^3} \frac{(1-x)^2}{x^2} + \order(\beta^2).
\end{equation}
Substitution into \cref{eq:fRpotential} yields
\begin{equation}
U(\varphi) = \frac{1}{8\alpha}(1-x)^2 \left[ 1 - \frac{\beta}{2\alpha^2} \frac{1-x}{x} + \order(\beta^2) \right].
\end{equation}
The enhancement factor \((1-x)/x\simeq4N/3\) on the plateau gives the control condition
\begin{equation}
\left|\frac{\beta}{\alpha^2}\right|\frac{1}{x_*}\ll1, \qquad \hbox{or approximately}\qquad \left|\frac{\beta}{\alpha^2}\right|N_*\ll1.
\label{eq:r3control}
\end{equation}
This comparison is included only to separate ordinary \(f(R)\) plateau deformations from the dynamical scalar--Gauss--Bonnet threshold used in the main text. In the \(R^3\) case the correction is an off-shell \(f(R)\) deformation and is enhanced by \(1/x_*\sim N_*\), forcing the control condition in \cref{eq:r3control}. In the scalar--Gauss--Bonnet case, by contrast, the constant Gauss--Bonnet part is topological and the observable deformation is extracted only after solving the stabilized threshold trajectory.

\section{Local extraction of \texorpdfstring{\(\kappa_G\)}{kappaG} from matched plateau data}
\label{app:kappaExtraction}
This appendix states how \(\kappa_G\) is extracted once an analytic or numerical single-clock background has been obtained. It is not a compactification construction. Given a solved adiabatic trajectory \(\psi_0(\varphi)\), fit the large-field background over the CMB window to
\begin{equation}
U_{\eff}(\varphi) = \Lambda^4\left[1+u_1x+u_2x^2+\order(x^3)\right], \qquad x=e^{-\sqrt{2/3}\varphi}.
\end{equation}
The leading falloff is normalized by
\begin{equation}
\tilde x=-\frac{u_1}{2}x,
\end{equation}
and the invariant extraction is
\begin{equation}
A_G=\frac{4u_2}{u_1^2}, \qquad \kappa_G=\frac32(A_G-1).
\end{equation}
The coefficients \(u_1,u_2\) are matched to background data. They should not be interpreted as coefficients of an off-shell potential obtained by replacing \(\GB\) with its background value. Operationally, the fit should be performed only over the asymptotic window relevant for the CMB pivot scales, not over the entire field range down to the end of inflation. The reason is that the expansion parameter \(x=e^{-\sqrt{2/3}\varphi}\) is small in the plateau regime, whereas the end-of-inflation region can receive corrections that are irrelevant for the leading attractor coefficients. In a numerical implementation, one should therefore extract \(u_1\) and \(u_2\) from the accepted single-clock background over the CMB window and then verify that the neglected \(\order(x^3)\) terms do not affect \(\ns\), \(\rten\), and \(\alphas\) at the quoted order.

This extraction can be applied to any analytic or numerical solution for the accepted adiabatic trajectory. The trajectory may come from a stabilized heavy-field equation, but that equation is only a way of determining \(\psi_0(\varphi)\). The scalar--Gauss--Bonnet term should not be promoted to an ordinary off-shell scalar potential by replacing \(\GB\) with its FLRW value. The same solved background must also satisfy the single-clock, weak-\(\delta_{\rm GB}\), stability, and cutoff conditions in \cref{eq:smallDeltaGB,eq:eftHierarchy}. Outside this weak-coupling regime, or for precision spectra, the extraction must be replaced by the full scalar--Gauss--Bonnet quadratic action.

Thus, the appendix defines the invariant background deformation used as a weak-coupling diagnostic; in a full numerical implementation, \(u_1\) and \(u_2\) should be extracted directly from the solved background.

\bibliography{string_corrected_horndeski_starobinsky}

\end{document}